\documentclass[12pt]{article}

\usepackage{amsmath,amsfonts,amssymb}
\usepackage{graphicx,color}
\usepackage{psfrag}
\usepackage{enumerate}
\usepackage{mathrsfs}

\makeatletter \@addtoreset{equation}{section}

\makeatletter\renewcommand\section{\@startsection {section}{1}{\z@}%
                                   {-3.5ex \@plus -1ex \@minus -.2ex}
                                   {2.3ex \@plus.2ex}%
                                   {\normalfont\large\bfseries}}
\renewcommand\subsection{\@startsection{subsection}{2}{\z@}%
                                     {-3.25ex\@plus -1ex \@minus -.2ex}%
                                     {1.5ex \@plus .2ex}%
                                     {\normalfont\bfseries}}

\newcommand{\be}{\begin{equation}}
\newcommand{\ee}{\end{equation}}

\newcommand{\eeq}{\end{eqnarray}}

\def\[{\left [}
\def\]{\right ]}
\def\({\left (}
\def\){\right )}

\def\r2{\sqrt{2}}

\newcommand{\bbibitem}[1]{\bibitem{#1}\marginpar{#1}}

\def\Label#1{\label{#1}%
  \smash{\hbox to0pt{\raise1ex\hbox{\tiny[#1]}\hss}}}
\def\noLabels{\let\Label=\label}
\def\nobbibitem{\let\bbibitem=\bibitem}

\begin{document}
\noLabels 
\nobbibitem 




\begin{center}

\begin{flushright} \vspace{-3cm}
{\small
UPR-1222-T}  \\ 
\end{flushright}
\vspace{2cm}

{\Large \bf Thin walls and junctions: Vacuum decay 
\\ \vspace{2mm} in multidimensional field landscapes}

\vspace{3mm}

Vijay Balasubramanian$^{a,}$\footnote{\tt email:vijay@physics.upenn.edu}, 
Bart{\l}omiej Czech, Klaus Larjo, Thomas S. Levi$^{b,}$\footnote{\tt email:czech, larjo, tslevi@phas.ubc.ca}

\vspace{5mm}

\bigskip\centerline{$^a$\it Department of Physics and
Astronomy}
\smallskip\centerline{\it David Rittenhouse Laboratories}
\smallskip\centerline{\it University of Pennsylvania}
\smallskip\centerline{\it 209 S 33$^{\rm rd}$ Street, Philadelphia, PA 19104, USA}
\bigskip\medskip
\bigskip\centerline{$^b$\it Department of Physics and
Astronomy}
\smallskip\centerline{\it University of British Columbia}
\smallskip\centerline{\it 6224 Agricultural Road, Vancouver,
BC V6T 1Z1, Canada}
\bigskip\medskip


\end{center}
\begin{abstract}
\noindent
We study tunneling between vacua in multidimensional field spaces in the thin wall approximation. The thin wall action generically gives rise to two types of saddle points. The first type corresponds to the well-known spherical instantons, but when the field space is multidimensional we find that it is necessary to include codimension two  ``junctions'' to regulate and compute the fluctuation determinant and the tunneling rate.   The second type of saddle point is novel and is present only when the field space is more than one-dimensional. These saddle points have two or more negative modes, which confirms that in the presence of additional field directions spherical instantons continue to dominate vacuum decay. We identify regimes in parameter space in which additional field directions may quench certain channels of vacuum decay.
\end{abstract}

\addtocounter{footnote}{-2}
\newpage

\section{Introduction}
In classic papers Callan and Coleman \cite{Coleman:1977py, Callan:1977pt, colemanbook} explained how to compute the rate of tunnelling between vacua in field space using the method of instantons.  The final answer takes the form
\begin{equation}
\Gamma \approx e^{-S_0} \left| \det \delta^2 S\right|^{-1/2}, \label{usual}
\end{equation}
where $S_0$ is the Euclidean action of the instanton mediating the transition. An important ingredient in obtaining an analytic expression is the thin wall approximation, which represents the instanton as a thin, spherical, tensionful wall, whose interior is filled with the true vacuum. The spherical form of the minimal action instanton was rigorously established in \cite{colemanproof}.

This paper extends the above analyses to the case where the field space is at least two-dimensional and contains at least three minima. We construct all critical points of the thin wall action and find that the action generically has two critical points.   
One of these represents the usual spherical instanton of a single bubble nucleating in the vacuum.  In the na\"\i ve thin wall limit, the fluctuation determinant around this saddle point has a vanishing eigenvalue, but this is regulated by the contributions of codimension two ``junctions'' where bubbles of three vacua meet.  The second class of critical points is novel. We show that these critical points have at least two negative modes and, consequently, do not mediate decays.  Thus, our analysis confirms that the standard thin wall analysis is robust to the addition of extra field directions in the first approximation and the thin wall limit.    These computations comprise Sec.~\ref{calculation}. In Sec.~\ref{discussion} we discuss the validity of our calculations and identify regimes in parameter space in which additional field directions may destabilize the spherical instantons.

\section{Bubbles in multidimensional field spaces}
\label{calculation}

Consider a two-dimensional potential landscape with three vacua as depicted in Fig.~\ref{landscape}. A universe filled with the false vacuum A is subject to tunneling events, which spontaneously produce non-trivial field profiles.  Let us assume that all relevant tunneling events can be analyzed in the thin wall approximation. Two necessary conditions for this are that the vacua of the scalar potential are nearly degenerate, and that the walls of the potential are high compared to the difference in vacuum energies. We implement these conditions by starting with a potential $V_0(\vec{\phi})$, for which the three vacua are degenerate, and then add a small degeneracy breaking term $\vec{\epsilon} \cdot \vec{\phi}$:
\begin{eqnarray}
& & V(\vec{\phi})  =  V_0(\vec{\phi}) + a^{-1}\vec{\epsilon} \cdot \vec{\phi}, \quad \textrm{with } V_0(\vec{\phi}_A)=V_0(\vec{\phi}_B)=V_0(\vec{\phi}_C) \equiv V_0,  \,\, \textrm{and}\Label{tw1} \\
& & V_0(\vec{\phi})-V_0   \gg  |\epsilon|, \quad \textrm{when $\vec{\phi}$ is not near any of the  vacua A, B or C.} \Label{tw2}
\end{eqnarray}
Here $a$ is a length scale in field space. To leading order in $\vec{\epsilon}$ the locations of the vacua are unaffected by the degeneracy breaking term.  We denote the energy density differences of the three vacua by
\begin{equation}
V(\vec{\phi}_A) - V(\vec{\phi}_B) \equiv \epsilon_{AB}, \quad V(\vec{\phi}_A) - V(\vec{\phi}_C) \equiv \epsilon_{AC}, \quad  V(\vec{\phi}_B) - V(\vec{\phi}_C) \equiv \epsilon_{BC} \Label{epsilons}
\end{equation}
with all $\epsilon$'s positive. Without loss of generality we have chosen A to be the false vacuum, B the intermediate vacuum, and C the true vacuum. In this approximation a tunneling event produces regions of vacua B and/or C, separated from the ambient vacuum A and from one another by thin walls, whose tensions are given by \cite{negeig}
\begin{eqnarray}
\sigma_{AB} & = & \min_l 
\int_A^B dl\, \sqrt{2 (V_0(\vec{\phi}_l) - V_0)}, \label{tensiondef} \\
\sigma_{AC} & = & \min_l 
\int_A^C dl\, \sqrt{2 (V_0(\vec{\phi}_l) - V_0)}, \label{tensiondefc} \\
\sigma_{BC} & = & \min_l 
\int_B^C dl\, \sqrt{2 (V_0(\vec{\phi}_l) - (V_0-\epsilon_{AB})} \approx \min_l
\int_B^C dl\, \sqrt{2 (V_0(\vec{\phi}_l) - V_0)} \label{tensiondeff},
\end{eqnarray}
where $l$ ranges over paths in field space that interpolate between the respective pairs of vacua.  
The definitions imply triangle inequalities:
\begin{eqnarray}
\sigma_{AC} & < & \sigma_{AB} + \sigma_{BC} \nonumber \\ 
\sigma_{AB} & < & \sigma_{AC} + \sigma_{BC} \label{trineq} \\
\sigma_{BC} & < & \sigma_{AB} + \sigma_{AC} \nonumber
\end{eqnarray}

\begin{figure}[t]
\begin{center}
$\begin{array}{cc}
\includegraphics[scale=0.75]{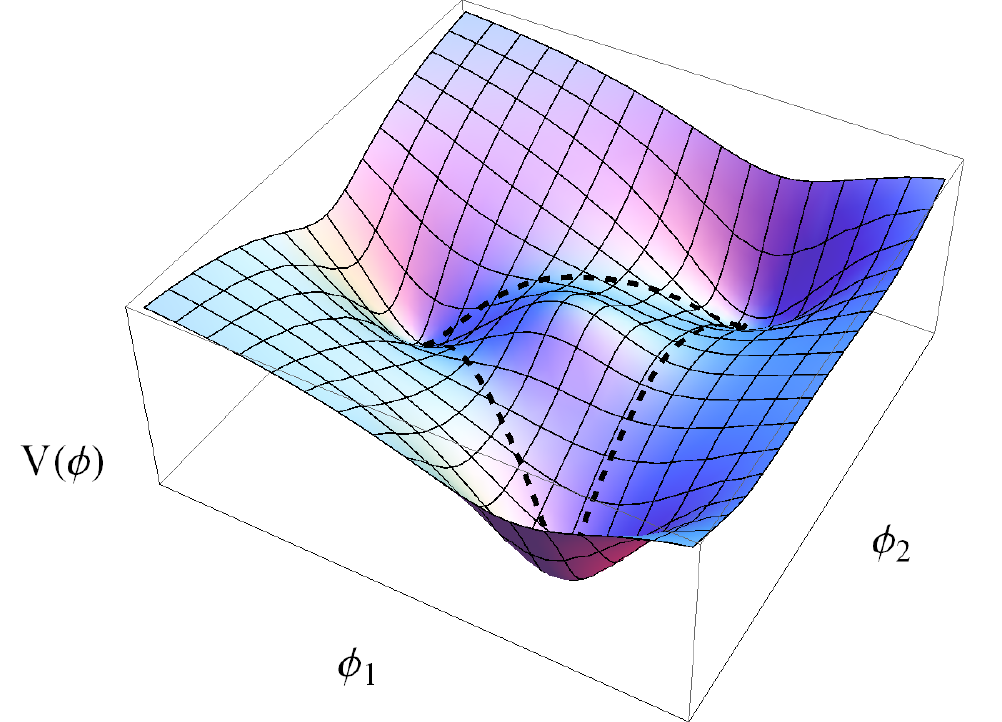}\qquad & 
\includegraphics[scale=0.35]{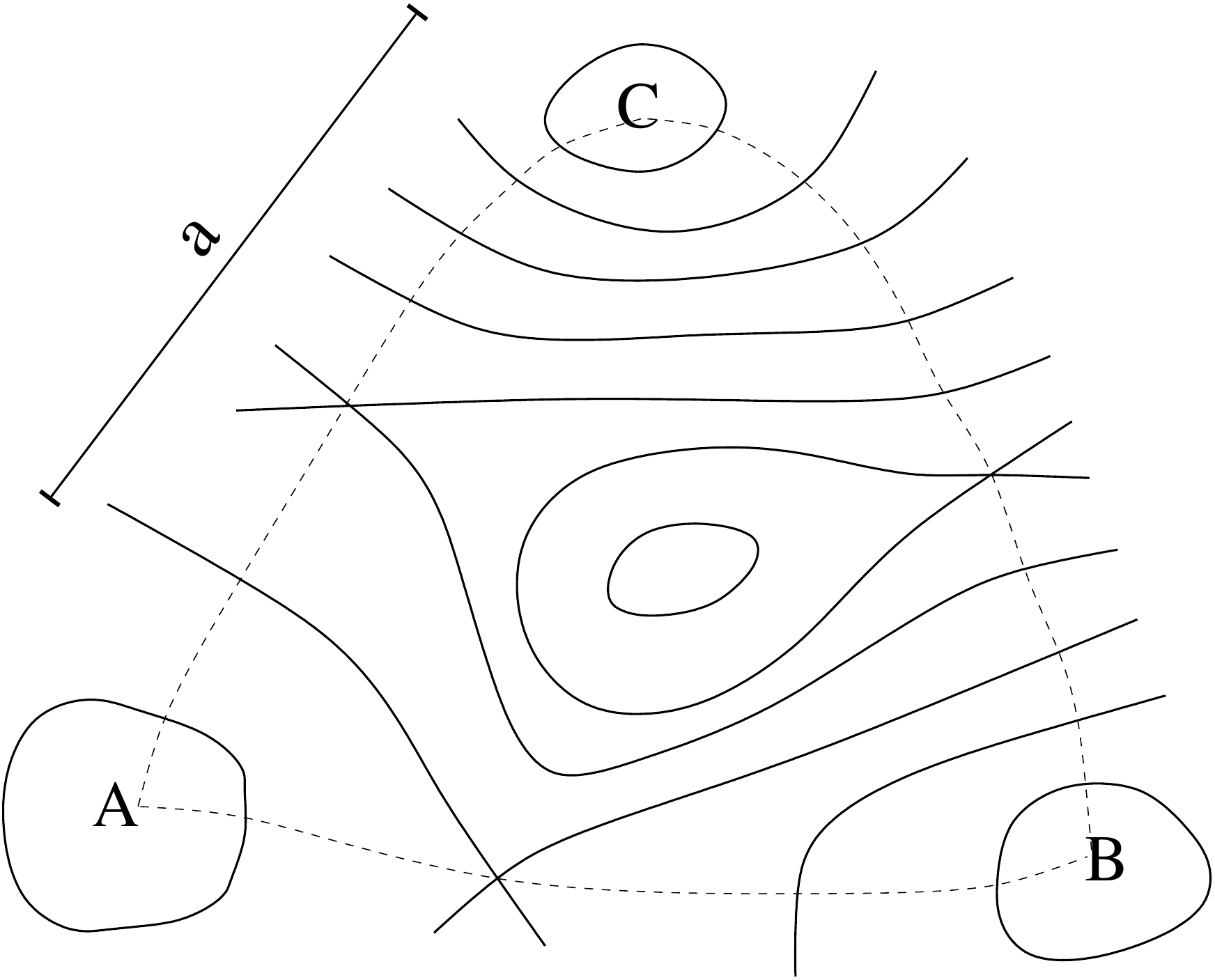}
\end{array}$
\caption{Left: A generic potential in the landscape with three vacua. Right: A contour plot of the same potential.   Dashed lines are minimal interpolating paths, which determine the tensions of the walls that in the thin wall approximation demarcate regions filled with the different vacua.}
\Label{landscape}
\end{center}
\end{figure}

\subsection{Thin wall configurations}

Fig.~\ref{general} illustrates a general configuration that may be assembled from the ingredients available in the thin wall approximation. However, it is sufficient to consider connected configurations, because each component of a disconnected configuration nucleates independently of the others. Further, it suffices to study connected configurations that consist of a single B-region and a single C-region, screened by a single $\sigma_{BC}$-interface. If we could find such a solution, we could then glue together fragments of it to arrive at more complicated configurations with many B- and C-regions separated by many interfaces. The current section concentrates on configurations built up of one B-region and one C-region, separated by a single interface.

\begin{figure}[t]
\begin{center}
\includegraphics[scale=0.4]{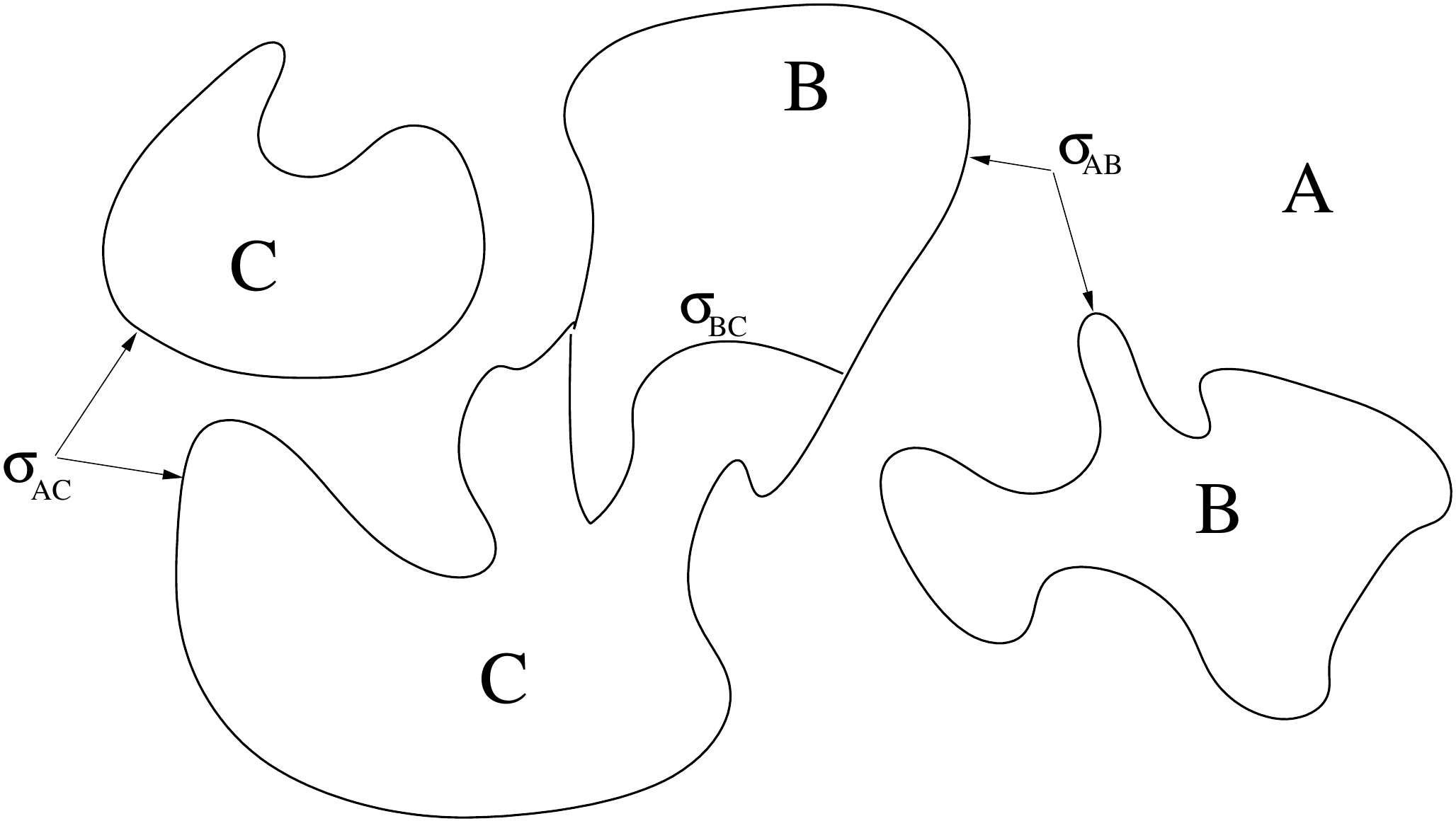}
\caption{A general configuration that can be assembled from thin wall ingredients.}
\Label{general}
\end{center}
\end{figure}

A configuration of this type contains a preferred spacelike axis, which joins the centers of mass of regions B and C. This special axis breaks down the symmetry of the problem from $SO(3,1)$ down to $SO(2,1)$. The lowest action instanton is expected to obey the symmetry. Hence, we will assume that the locus where regions A, B, C coalesce is a two-hyperboloid. We shall refer to this locus as the junction. On the time slice of nucleation, which we pick to be $t=0$, the junction forms a circle; we shall denote its radius $r$. The walls separating the different regions are now subject to a boundary condition: each of them has a boundary given by a circle of radius $r$ (at $t=0$) or a two-hyperboloid (including the time direction).   Without this new boundary condition, the instanton minimizing the action would describe a wall that is a two-sphere at $t=0$ \cite{colemanproof}\label{so4arg}. A sliced sphere will thus also minimize the action while obeying the desired boundary condition of having a circular boundary. In the end, we see that the walls of our bubbles must be sliced two-spheres (at time $t=0$) or three-hyperboloids (including the time direction).

\begin{figure}[t]
\begin{center}
\includegraphics[scale=0.4]{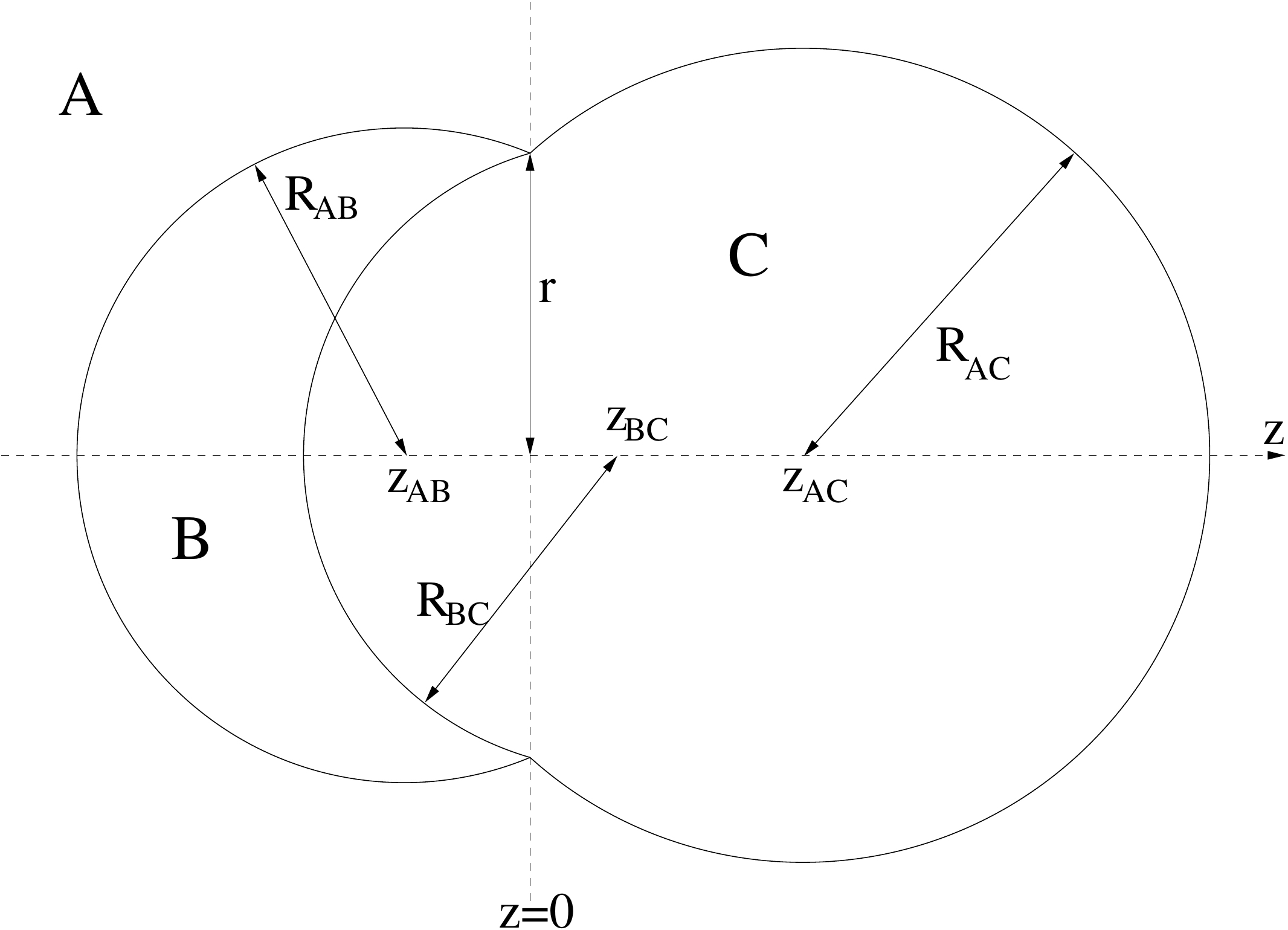}
\caption{Our ansatz for the initial ($t=0$) configuration that nucleates. The coordinate $z$ increases to the right.}
\Label{setup}
\end{center}
\end{figure}

Based on this argument, we may restrict attention to configurations that at time $t=0$ resemble Fig.~\ref{setup}. The walls trace two-spheres, which meet on a circle of radius $r$. We denote the radii of the spheres by $R_{AB}, R_{AC}, R_{BC}$, where $R_{AB}$ ($R_{AC}$) screens the region filled with vacuum B (C) from the ambient vacuum A while $R_{BC}$ is the radius of the spherical surface separating bubbles B and C. We will refer to these (partial) spheres as AB, AC and BC. Note that since C has a lower vacuum energy than B, the spherical surface defined by $R_{BC}$ will always bend into bubble B. Because the three spheres meet on a common circle, their centers are colinear. The axis joining their centers, which breaks the initial $SO(3,1)$ symmetry down to $SO(2,1)$, will be denoted $z$. We set $z=0$ to mark the plane containing the $r$-circle common to the three spheres. The centers of the three spherical surfaces are then marked $z_{AB},\, z_{AC},\, z_{BC}$.  Although Fig.~\ref{setup} shows a particular choice of $z_{AB},\, z_{AC},\, z_{BC}$, the analysis below is general and makes no assumptions about their signs.

\begin{figure}[t]
\begin{center}
\includegraphics[scale=0.35]{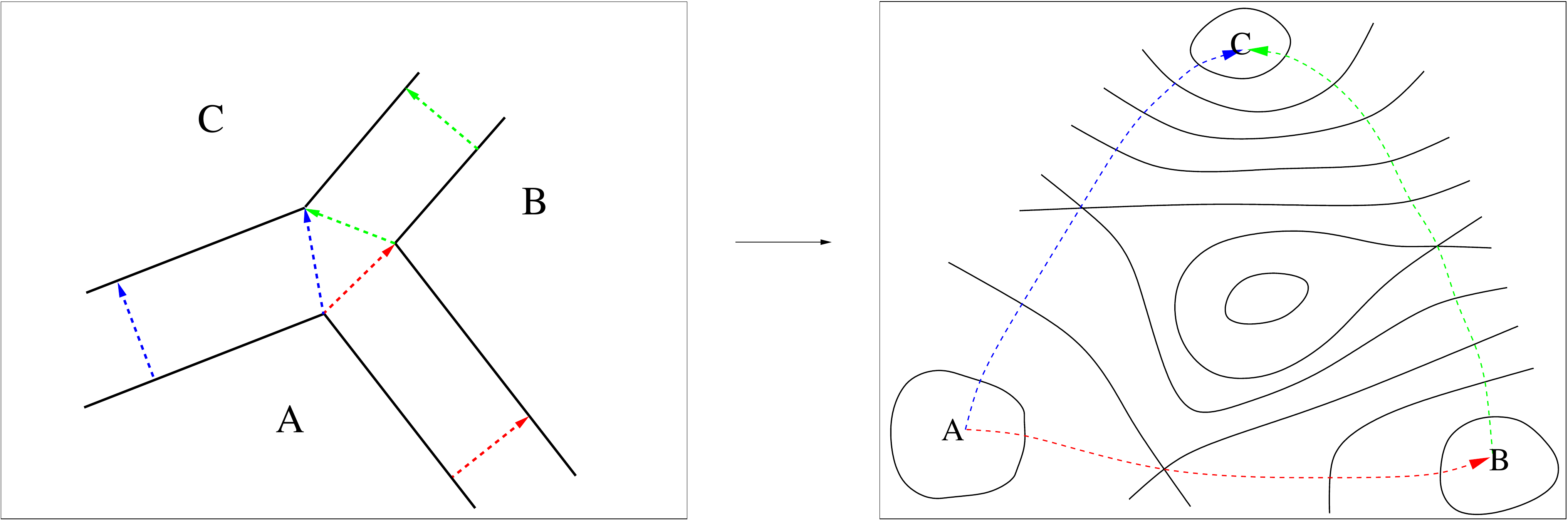}
\caption{The field profile over the junction. It is the image of a map from the cross section of the junction (a triangle in space at any fixed time) to the region in field space lying between the three vacua.}
\Label{junction}
\end{center}
\end{figure}

\paragraph{Junctions:} The above discussion assumed that we are free to join different types of bubble walls on junctions. This requires the existence of an appropriate solution to the field equations of motion in the interior of the junction. To consider what such a solution should look like, consult Fig.~\ref{junction}.  The boundary conditions for $\vec{\phi}$ on the junction are that the sides of the junction-triangle in Euclidean space map to the minimal paths defining $\sigma_{AB},\,\sigma_{AC},\,\sigma_{BC}$  in field space.  We believe that the requisite solution exists, though we have no formal proof. 

\subsection{The action}
\label{theaction}

To find the Euclidean action of an instanton, we integrate the Euclidean Lagrangian evaluated on Wick-rotated solutions to the equations of motion, subtracting the action of the ambient vacuum \cite{colemanbook}. In the case of the spherical instanton, the Wick rotation turns $SO(3,1)$-invariant hyperboloids into $SO(4)$-invariant spheres. Thus, the imaginary part of the Euclidean action for a spherical bubble of radius $R$ takes the form
\begin{equation}
S \!=\! -\epsilon R^4 {\rm Vol} (B_4) + \sigma R^3 {\rm Vol}(S^3).
\label{cdlorig}
\end{equation}
In the language of \cite{colemanbook}, this is the action of a \emph{full} bounce, not a half-bounce, as is preferred by some authors, e.g. \cite{esko}. Thus, the nucleation rate depends on $S$ according to $\Gamma \propto \exp{(-S)}$.

We would like to write down an analogue of equation~(\ref{cdlorig}) for the configuration presented in Fig.~\ref{setup}. It respects $SO(2,1)$ symmetry, which is $SO(3)$ in the Euclidean section. Thus, after we restore the dependence on Euclidean time, the slices of two-spheres on each side of the plane $z=0$ become slices of $S^3$'s. As a consequence of the $SO(3)$ symmetry of the Euclidean solution, after the Wick rotation all the three-spheres comprising the solution continue to meet on the same two-sphere. In the end, the correct action consists of three contributions from the slices of spheres AB, AC and BC, each of which is of the form (\ref{cdlorig}), except that the volumes of $B_4$ and $S^3$ are replaced by  the volumes of their respective slices.

To write down this action, we need the volumes of slices of $B_4$ and $S^3$. Consider the unit ball $B_4$, surrounded by a unit $S^3$, centered at the origin and intersected by a plane $z=s$. This plane splits the sphere into two parts,  the `left' part whose points satisfy $-1\le z\le s$, and the `right' part whose coordinates satisfy $s\le z\le 1$.  The volume of $B_4$ and the three-dimensional area of $S^3$ of the left part are given by
\begin{eqnarray}
a(s) & \equiv & \int_0^{2\pi} d\phi \int_0^\pi d\theta \sin\theta \int_{\cos^{-1}\!{s}}^\pi \!d\psi\, \sin^2\psi =  \pi \left( \pi + 2\sin^{-1}{s} + 2s \sqrt{1-s^2} \right), \label{s3}\\
v(s) & \equiv & \int_{-1}^s \!dz\, \frac{4\pi}{3} \left( 1-z^2\right)^{3/2} = \frac{\pi^2}{4} + \frac{\pi}{6} \left( s (5 - 2 s^2) \sqrt{1-s^2} + 3 \sin^{-1}s\right), \label{b4}
\end{eqnarray}
where we have chosen the branch  $\sin^{-1} s \in [-\pi/2,\pi/2]$. Note that due to the reflection symmetry $(z\to -z)$, the area and volume of the right part are given by $a(-s)$ and $v(-s)$.  Using these definitions, contributions of each sliced sphere contain factors of the form $a(\pm z_X/R_X)$ and $v(\pm z_X/R_X)$. Further, from the geometry (see Fig.~\ref{setup}) we find 
\begin{equation}
r^2 + z_X^2= R_{X}^2, \quad \textrm{ with } X = AB, \, AC, \, BC. \label{rzrelations}
\end{equation}
Using these identities, we may write the contribution  $S^X$ of sphere X to the total action $S$ in terms of $r$ and the radii $R_X$ as
\begin{equation}
S^X = - \epsilon_X R_X^4 \,v\!\left(\pm \sqrt{1-\frac{r^2}{R_X^2}}\right) + \sigma_X R_X^3 \,a\!\left(\pm \sqrt{1-\frac{r^2}{R_X^2}}\right), \label{sfunctionalr}
\end{equation}
where the upper sign is chosen if the larger slice of the sphere is retained, and the lower sign is chosen if the smaller slice is kept.  The easiest way to see this is that $a(s)$ and $v(s)$ are monotonically increasing functions, and the larger the retained part of the sphere, the larger the contribution to the action must to be. The action for the full configuration is then simply
\begin{equation}
S = S^{AB} + S^{AC} + S^{BC}. \label{sfunctional}
\end{equation}
Note that the volume of the slice of sphere BC is counted twice, with prefactors $\epsilon_{AB}$  and $\epsilon_{BC}$.  However, the vacuum inside this  sphere is C, and hence the prefactor has  to be $\epsilon_{AC}$.  Using (\ref{epsilons}) we find $\epsilon_{AB}+\epsilon_{BC} = \epsilon_{AC}$, so the prefactor works out just right.  Equation~(\ref{sfunctional}) generalizes action (\ref{cdlorig}) for intersecting bubbles. The four parameters $r$, $R_{AB}$, $R_{AC}$, $R_{BC}$ are found by extremizing (\ref{sfunctional}).

\paragraph{Junctions:} The Euclidean action receives an additional contribution from the junction locus. This term is only present when there are two or more dimensions in field space. To estimate the magnitude of this contribution, think of the three bubble walls meeting at the junction as shells of thickness $\mu$, which are (slices of) three-dimensional spheres. Fig.~\ref{junction} shows that the junction can be roughly thought of as an intersection of such shells. Thus, the junction is itself a shell of cross section $\mu^2$ and the topology of a two-sphere. The radius of the $S^2$ is $r+\mu$, as can be seen from Figs.~\ref{setup} and \ref{junction}. Overall, the junction contributes to the action a term of the form $4\pi \mu^2 (r+\mu)^2 \kappa \approx 4 \pi \kappa \mu^2 r^2$, where $\kappa$ is some quantity that depends on the scalar potential in the interior of the triangle ABC in field space (compare with the right panel in Fig.~\ref{junction}). 

We work in the thin wall approximation, which requires $\mu \ll R_X$ for $X={AB, AC, BC}$. We further asssume that $\kappa$ is not large enough to compensate for the smallness of $\mu$. Under these approximations, we can consistently ignore the junction term for finding the {\it locations} of the critical points. However, we will see that  junctions can play an important role in regulating the second variation of the action around the saddle point and thus are important for correctly computing the tunneling rate.

\subsection{Saddle points}
\label{extr}

We now extremize action (\ref{sfunctional}). We limit our attention to variations parameterized by $r$, $R_{AB}$, $R_{AC}$, $R_{BC}$, because as mentioned on p.~\pageref{so4arg}, all other variations increase the action. We shall see in the next section that using $r$, $R_{AB}$, $R_{AC}$, $R_{BC}$  to parameterize variations is the most judicious choice, because it diagonalizes the Hessian of (\ref{sfunctional}) at all critical points.

It is easy to extremize with respect to $R_X$. Because $S^{AC}$ and $S^{BC}$ are independent of $R_{AB}$, the extremization condition is simply
\begin{equation}
\frac{\partial S^{AB}}{\partial R_{AB}} = 0
\qquad \Longrightarrow \qquad
R_{AB}^* = \frac{3\sigma_{AB}}{\epsilon_{AB}}, \label{radiusextreme}
\end{equation}
and likewise for $R_{AC}$ and $R_{BC}$, independently of $r$. The extremal value  $R_{AB}^*$ agrees with the critical radius of the spherical bubble \cite{Coleman:1977py, colemanbook}. In consequence, any extremum of (\ref{sfunctional}) must have bubble radii set by (\ref{radiusextreme}).  

To extremize with respect to $r$, substitute (\ref{radiusextreme}) into eqs.~(\ref{sfunctionalr}-\ref{sfunctional}). Defining $k(s)$ as the difference of eq.~(\ref{s3}) and three times eq.~(\ref{b4}),
\begin{equation}
k(s) = \frac{\pi}{4} \left( \pi + 2 s (2s^2 - 1) \sqrt{1-s^2} +2 \sin^{-1}s \right), \label{defk}
\end{equation}
we obtain for $S^X$
\begin{equation}
S^X(r) = \frac{\epsilon_X {R^*_X}^4}{3} \,k\!\left( \pm \sqrt{1 - \frac{r^2}{{R^*_X}^2}} \right), \label{sxr}
\end{equation}
with the upper (lower) sign corresponding to cases where more (less) than half of bubble $X$ is retained. These two cases are distinguished by the signs of the variables $z_X$, which affords a general expression for $S(r)$:
\begin{multline}
S(r) = \frac{\epsilon_{AB} {R^*_{AB}}^4}{3} \,k\!\left( -{\rm sign}(z_{AB}) \sqrt{1 - \frac{r^2}{{R^*_{AB}}^2}}\right) +\frac{\epsilon_{AC} {R^*_{AC}}^4}{3} \,k\!\left( +{\rm sign}(z_{AC}) \sqrt{1 - \frac{r^2}{{R^*_{AC}}^2}}\right) \\+\frac{\epsilon_{BC} {R^*_{BC}}^4}{3} \,k\!\left( -{\rm sign}(z_{BC}) \sqrt{1 - \frac{r^2}{{R^*_{BC}}^2}} \right)
\label{sr}
\end{multline}
Using $dk(s)/ds = 4\pi s^2 \sqrt{1-s^2}$ and eq.~(\ref{rzrelations}) we obtain:
\begin{multline}
\frac{3}{4\pi} \frac{dS(r)}{dr} = \\ r^2 \left( \epsilon_{AB}\, {\rm sign}(z_{AB}) \sqrt{{R_{AB}^*}^2 - r^2} - \epsilon_{AC}\, {\rm sign}(z_{AC}) \sqrt{{R_{AC}^*}^2 - r^2} + \epsilon_{BC}\, {\rm sign}(z_{BC}) \sqrt{{R_{BC}^*}^2 - r^2} \right) \\
=  r^2 \left( \epsilon_{AB} z_{AB} - \epsilon_{AC} z_{AC} + \epsilon_{BC} z_{BC} \right) \label{firstder}
\end{multline}
There is one extremum at $r=0$ and one possible additional extremum that sets the quantity in parenthesis to zero.  Different choices of signs of $z_X$ pick different branches of $S(r)$ and represent distinct configurations. We discuss them in detail below. For each choice of signs of $z_X$, the extremum at $r=0$ corresponds to a familiar, spherical bubble or a combination thereof. The other extremum, if it exists, combines regions filled with vacua B and C. 

The quantity $r$ at a non-trivial extremum is therefore found by solving
\begin{equation}
\epsilon_{AB}\, {\rm sign}(z_{AB}) \sqrt{{R_{AB}^*}^2 - r^2} - \epsilon_{AC}\, {\rm sign}(z_{AC}) \sqrt{{R_{AC}^*}^2 - r^2} + \epsilon_{BC}\, {\rm sign}(z_{BC}) \sqrt{{R_{BC}^*}^2 - r^2} = 0.
\end{equation}
One may temporarily drop the information about the signs of $z_X$ by squaring twice. Using eq.~(\ref{radiusextreme}), this leads to:
\begin{eqnarray}
2 r^2 \big(\sigma_{AB}^2 (\epsilon_{AC}^2 +\epsilon_{BC}^2 - \epsilon_{AB}^2)+ \sigma_{AC}^2 (\epsilon_{BC}^2 +\epsilon_{AB}^2 - \epsilon_{AC}^2) +\sigma_{BC}^2 (\epsilon_{AB}^2 +\epsilon_{AC}^2 - \epsilon_{BC}^2)\big) & = & \nonumber \\ 
9 (\sigma_{AB} + \sigma_{AC} + \sigma_{BC}) (\sigma_{AB} + \sigma_{AC} - \sigma_{BC}) (\sigma_{AB} + \sigma_{BC} - \sigma_{AC}) (\sigma_{AC} + \sigma_{BC} - \sigma_{AB}) & = & 144 A_\triangle^2 \nonumber \\
& &
\label{masterr}
\end{eqnarray}
By Heron's formula, $A_\triangle$ is the area of the triangle whose sides are $\sigma_{AB}, \sigma_{AC}, \sigma_{AF}$, which exists by virtue of (\ref{trineq}). This yields at most one positive solution, call it $r_*$. 

The following argument shows that once $r_*$ is found, it picks a unique choice of signs for $z_X$. A priori, there are $2^3=8$ such choices, but two are outright excluded, because $z_{AC} < 0 < z_{AB}$ would prevent the BC-interface from ``fitting inside'' bubble B. Two others are excluded by conservation of energy. The choice $z_{AC} < 0 < z_{AB}, z_{BC}$ corresponds to taking ``less than half'' of the spheres AB, AC and BC, but such a configuration could not have a large enough interior to counter the energy present in the walls. Likewise, $z_{AB}, z_{BC} < 0 < z_{AC}$ corresponds to taking ``more than half'' of the three spheres, in which case the wall tension cannot balance the negative energy in the interior. Thus, one is left with four possible cases: 
\begin{equation}
\begin{array}{rcl}
z_{AB}, z_{AC}, z_{BC} < 0 & \qquad\qquad &  0 < z_{AB}, z_{AC}, z_{BC} \\
z_{AB}, z_{AC} < 0 < z_{BC} & \qquad\qquad & z_{AB} < 0 < z_{AC}, z_{BC} 
\end{array}
\label{casesz}
\end{equation}
Because $r_*$ is selected by solving
\begin{equation}
\epsilon_{AB} z_{AB} - \epsilon_{AC} z_{AC} + \epsilon_{BC} z_{BC}=0, \label{setrstar}
\end{equation}
we see that the only pair of cases (\ref{casesz}) which could potentially simultaneously solve (\ref{setrstar}) is the pair in the upper line. However, $z_{AB}, z_{AC}, z_{BC} < 0$ requires $R_{AB} > R_{BC}$ while $z_{AB}, z_{AC}, z_{BC} > 0$ demands $R_{AB} < R_{BC}$, or else the BC wall will not fit inside the B-bubble. Thus, the landscape parameters $\{\epsilon_{AB}, \epsilon_{AC}, \epsilon_{BC}, \sigma_{AB}, \sigma_{AC}, \sigma_{BC}\}$ uniquely set the locations of the three centers $z_X$. 

In summary, action (\ref{sfunctional}) contains two types of saddle points. Both have $R_X^*=3\sigma_X/\epsilon_X$, but they differ in their values of $r$. The saddle points at $r=0$ are (combinations of) spherical bubbles. For any given set of landscape parameters there may also exist at most one additional saddle point. It is characterized by $r=r_*$, the solution to (\ref{masterr}), supplemented by a choice of signs of $z_X$, which is uniquely selected by eq.~(\ref{setrstar}). 

\subsection{Eigenvalues}

We now compute the Hessian matrix of (\ref{sfunctional}) in order to extract the eigenvalues of the second variation of the action at the critical points. We begin with the off-diagonal terms. Because action (\ref{sfunctional}) is a sum of terms that depend on $R_{AB}$, $R_{AC}$ and $R_{BC}$ separately, 
\begin{equation}
S_{XY} =  0, \quad \textrm{with } X \neq Y, \quad X,Y \in \{AB,AC,BC\}.
\end{equation}
The remaining off-diagonal terms take the form
\begin{equation}
S_{rX} = \mp \frac{4\pi\epsilon_{X} r^4}{3(R_X^2 - r^2)^{3/2}}\left(R_X - \frac{3\sigma_X}{\epsilon_X} \right)
\end{equation}
(the signs correspond to eq.~\ref{sxr}) and vanish at all extrema by virtue of eq.~(\ref{radiusextreme}).  This means that when we are at a critical point, the Hessian is a diagonal matrix and its eigenvalues are $S_{rr}$ and $S_{XX}$. This reduces the counting of negative eigenvalues to four separate one-dimensional problems.

There is a simple criterion for the signs of $S_{XX}$. At the extremal value $R_{X}^*$ we find
\begin{equation}
S_{XX} \equiv \frac{\partial^2 S^{X}}{\partial R_{X}^2}\Big{|}_{R_{X}^* }\quad \left\{
\begin{array}{l}
< 0 \quad \textrm{if more than half of bubble X is retained} \\
> 0 \quad \textrm{if less than half of bubble X is retained.}
\end{array} \right. \label{sbbsign}
\end{equation}
Thus, each ``greater-than-half'' bubble slice entering a critical configuration comes with one negative eigenvalue, corresponding to the variation of its radius.

To determine the sign of $S_{rr}$, recall that in addition to the critical point at $r=0$, there is at most one other critical point. Thus, if one of them is a minimum of $S(r)$, the other must be a maximum and vice versa. Consequently, it is sufficient to expand eq.~(\ref{sr}) as a power series around $r=0$. The linear {and quadratic} terms vanish while the cubic coefficient is a positive multiple of (\ref{setrstar}). Since in the limit $r \to 0$ we have $z_X \to 3\, {\rm sign}(z_X)\, \sigma_X / \epsilon_X$, we obtain:
\begin{equation}
S(r) - S(0) = \frac{4\pi}{3}\big({\rm sign}(z_{AB})\, \sigma_{AB} - {\rm sign}(z_{AC})\, \sigma_{AC} + {\rm sign}(z_{BC})\, \sigma_{BC}\big) r^3 +\mathcal{O}(r^4) \label{taylor}
\end{equation}
Thus, the eigenvalue $S_{rr}$ vanishes at the $r=0$ critical points, although restoring the juction term $4 \pi \kappa \mu^2 r^2$ lifts this zero mode. If we continue to neglect junctions, the cubic term determines whether $S(r)$ is an increasing or decreasing function of $r$ between $0$ and $r_*$. Consequently, its sign allows us to identify $r_*$ as a minimum or maximum of $S(r)$, which fixes the sign of the eigenvalue $S_{rr}(r_*)$:
\begin{equation}
{\rm sign}(z_{AB})\, \sigma_{AB} - {\rm sign}(z_{AC})\, \sigma_{AC} + {\rm sign}(z_{BC})\, \sigma_{BC} > 0 \qquad \Longleftrightarrow \qquad S_{rr}(r_*) < 0
\end{equation}
This rule, supplemented by the triangle inequality (\ref{trineq}), determines the sign of $S_{rr}(r_*)$ in all cases (\ref{casesz}). The critical points of action (\ref{sfunctional}) are summarized in Table~\ref{allsols}.

\begin{table}[!t]
\begin{tabular}{|c|c|c|c|c|}
\hline
(1) & (2) & (3) & (4) & (5) \\
\hline
$(z_X)$ & $r=0$ & $r=r_*$ & eigenvalues at $r_*$ &  limit  $r_* \to 0$\\
\hline
$(+,-,+)$ &empty A-vacuum & \raisebox{-0.4in}[0.5in][0.5in]{\includegraphics[scale=0.2]{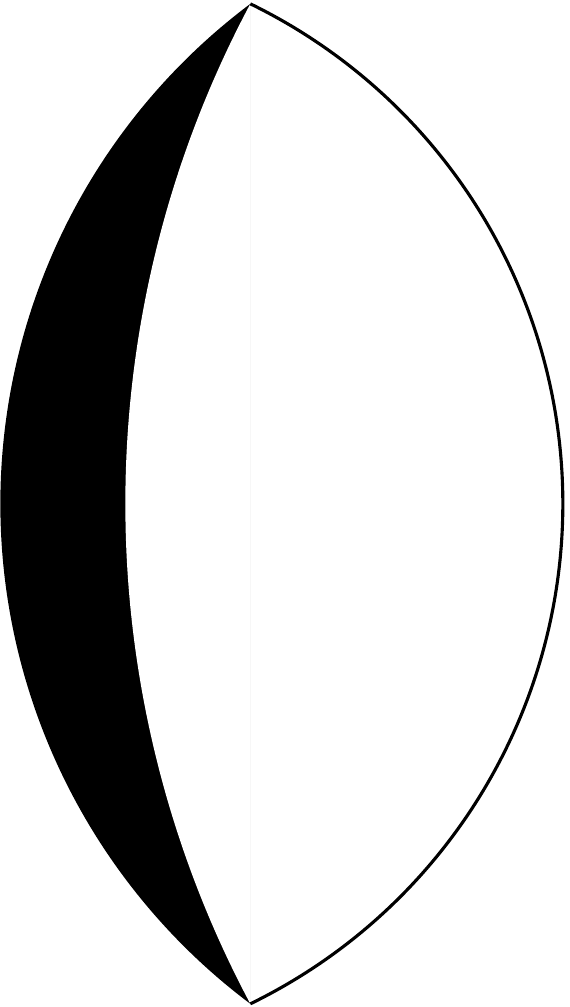}}  & $\times$   & \parbox{1.4in}{\begin{center}does not exist \\ ~ \\ excluded by energy conservation\end{center}} \\
\hline

$(+,+,+)$ &\raisebox{-0.3in}[0.4in][0.4in]{\includegraphics[scale=0.15]{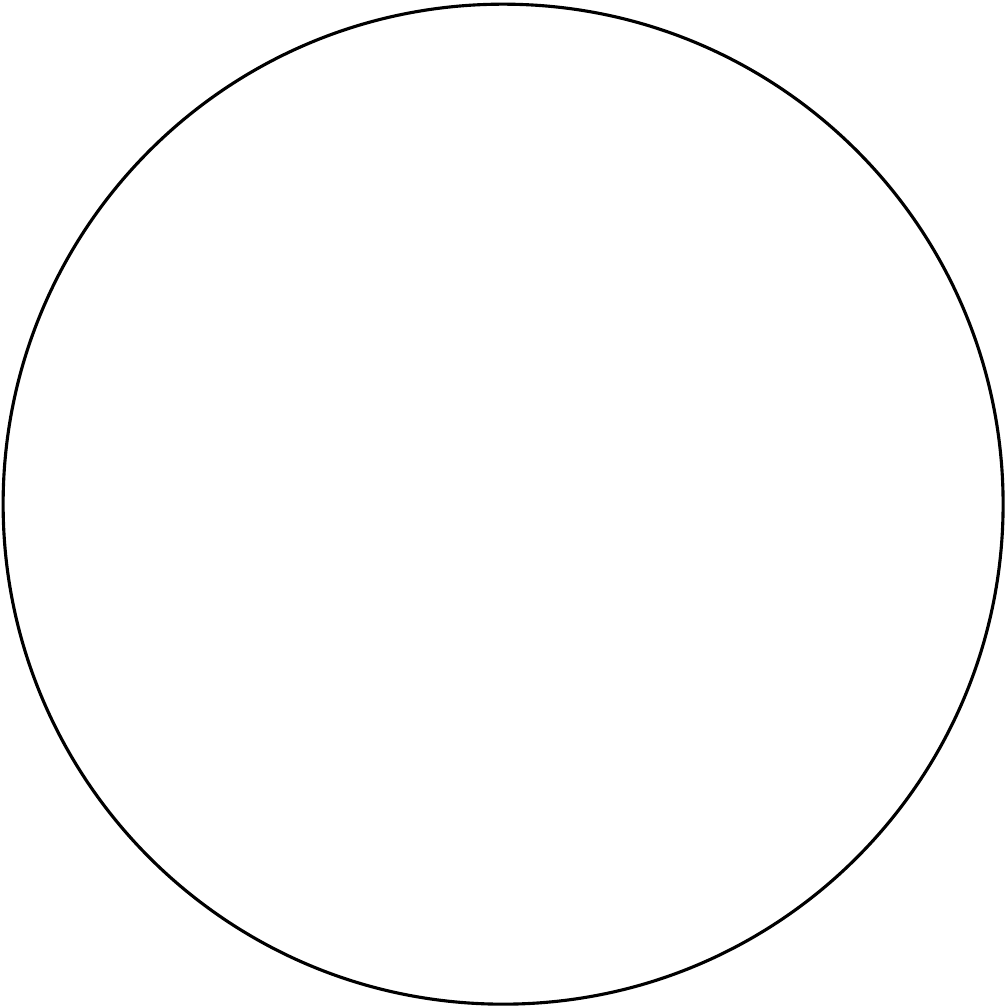}} &   \raisebox{-0.3in}[0.4in][0.4in]{\includegraphics[scale=0.2]{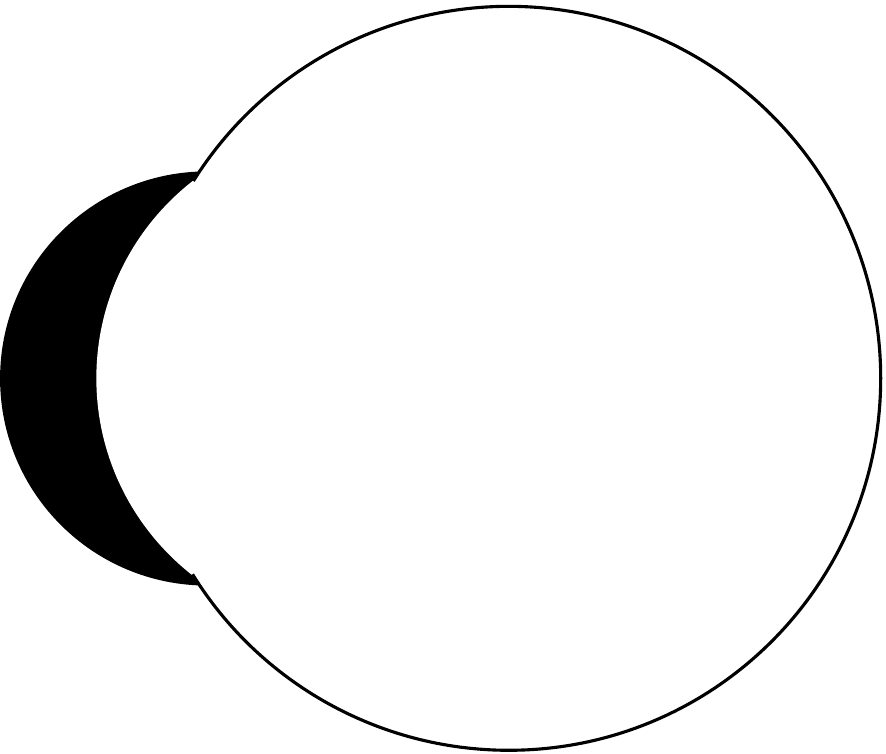}} 
 & $(-,+,-,+)$  & 
\parbox{1.4in}{\begin{center} $\sigma_{AC} \to \sigma_{AB} + \sigma_{BC}$ \\ 
\rule{0pt}{3ex} $\sigma_{AB} \, \epsilon_{BC} < \sigma_{BC} \, \epsilon_{AB}$ \end{center}} \\
\hline

$(-,-,-)$ & \raisebox{-0.4in}[0.5in][0.5in]{\includegraphics[scale=0.2]{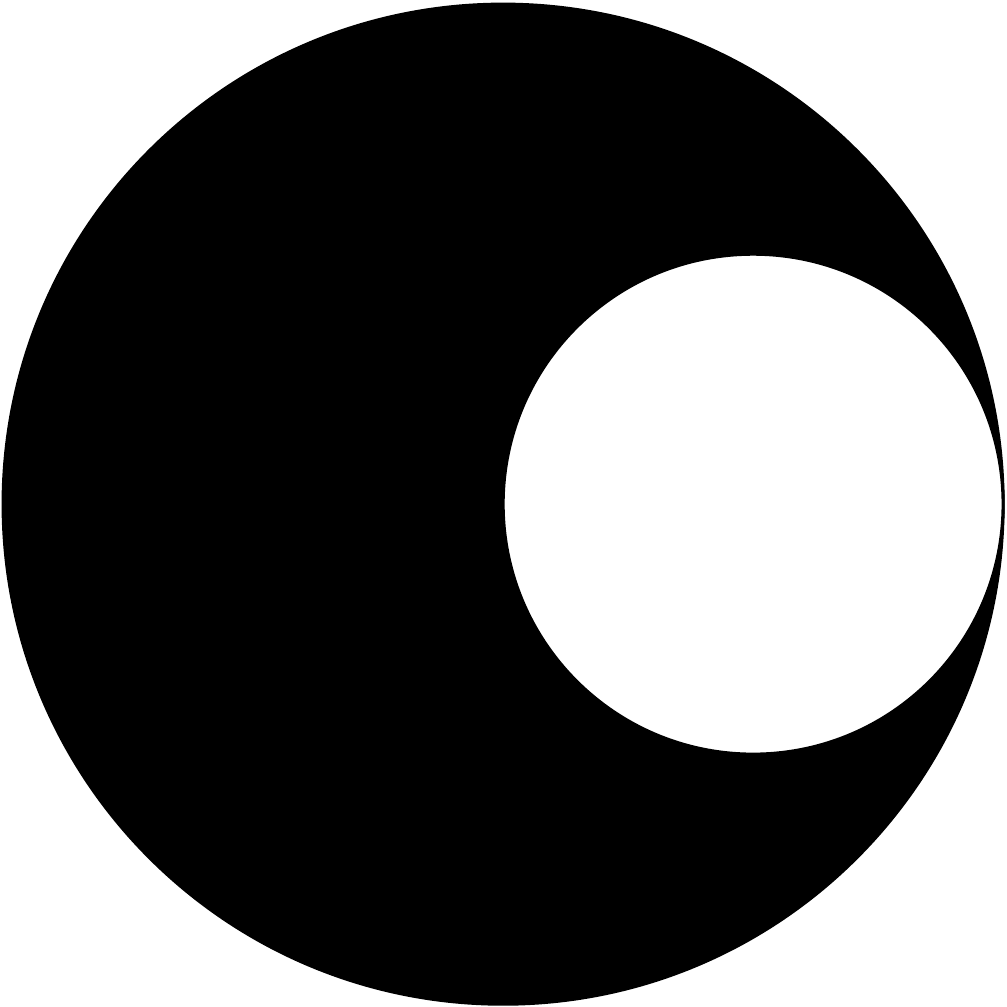}} & \raisebox{-0.4in}[0.5in][0.5in]{\includegraphics[scale=0.2]{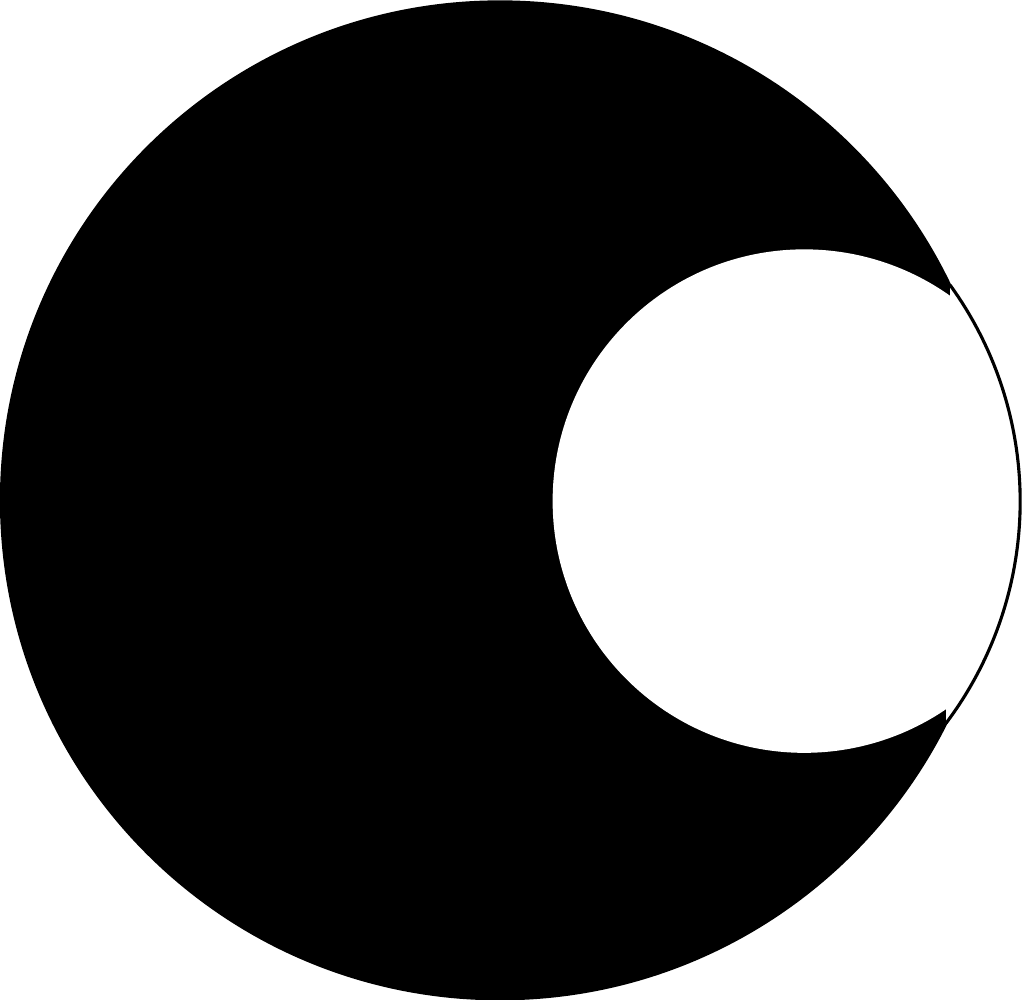}} &  $(+,-,+,-)$ &
\parbox{1.4in}{\begin{center} $\sigma_{AC} \to \sigma_{AB} + \sigma_{BC}$ \\ 
\rule{0pt}{3ex} $\sigma_{AB} \, \epsilon_{BC} > \sigma_{BC} \, \epsilon_{AB}$ \end{center}} \\
\hline

$(-,-,+)$ & \raisebox{-0.4in}[0.5in][0.5in]{\includegraphics[scale=0.2]{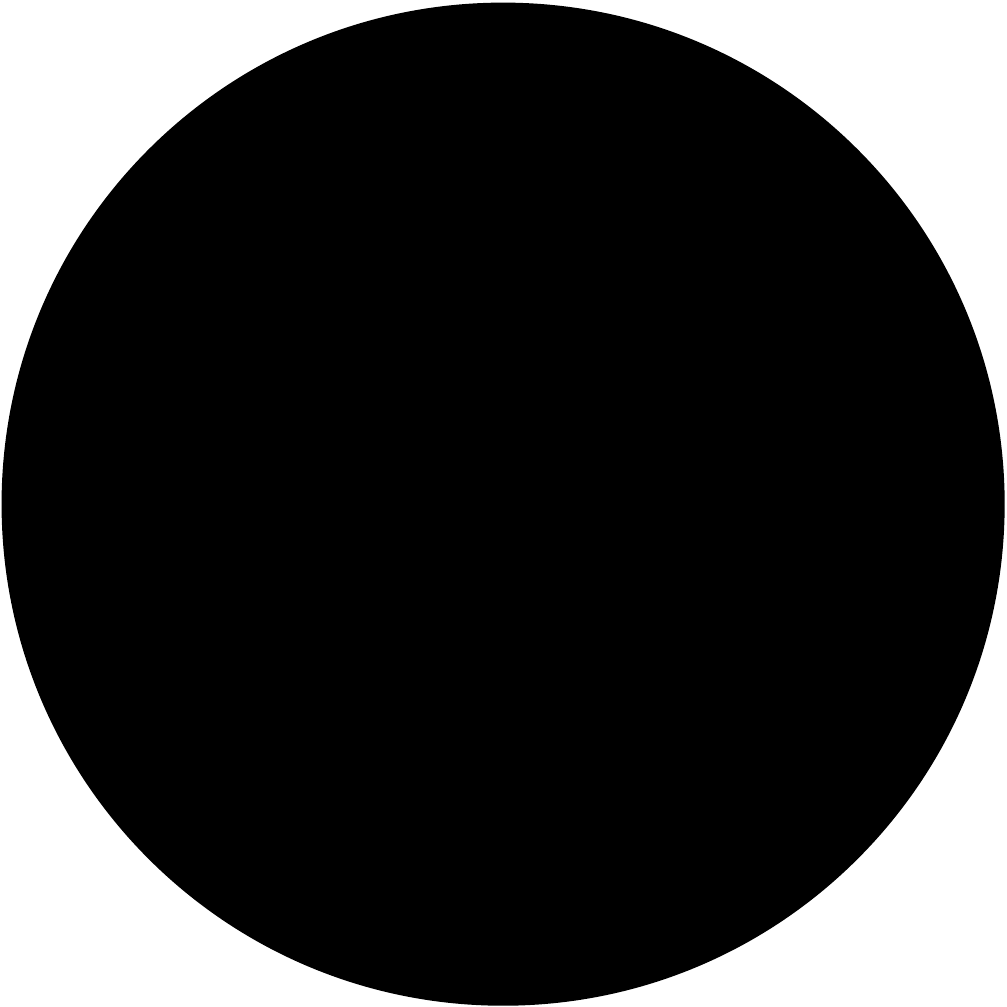}}  & \raisebox{-0.4in}[0.5in][0.5in]{\includegraphics[scale=0.2]{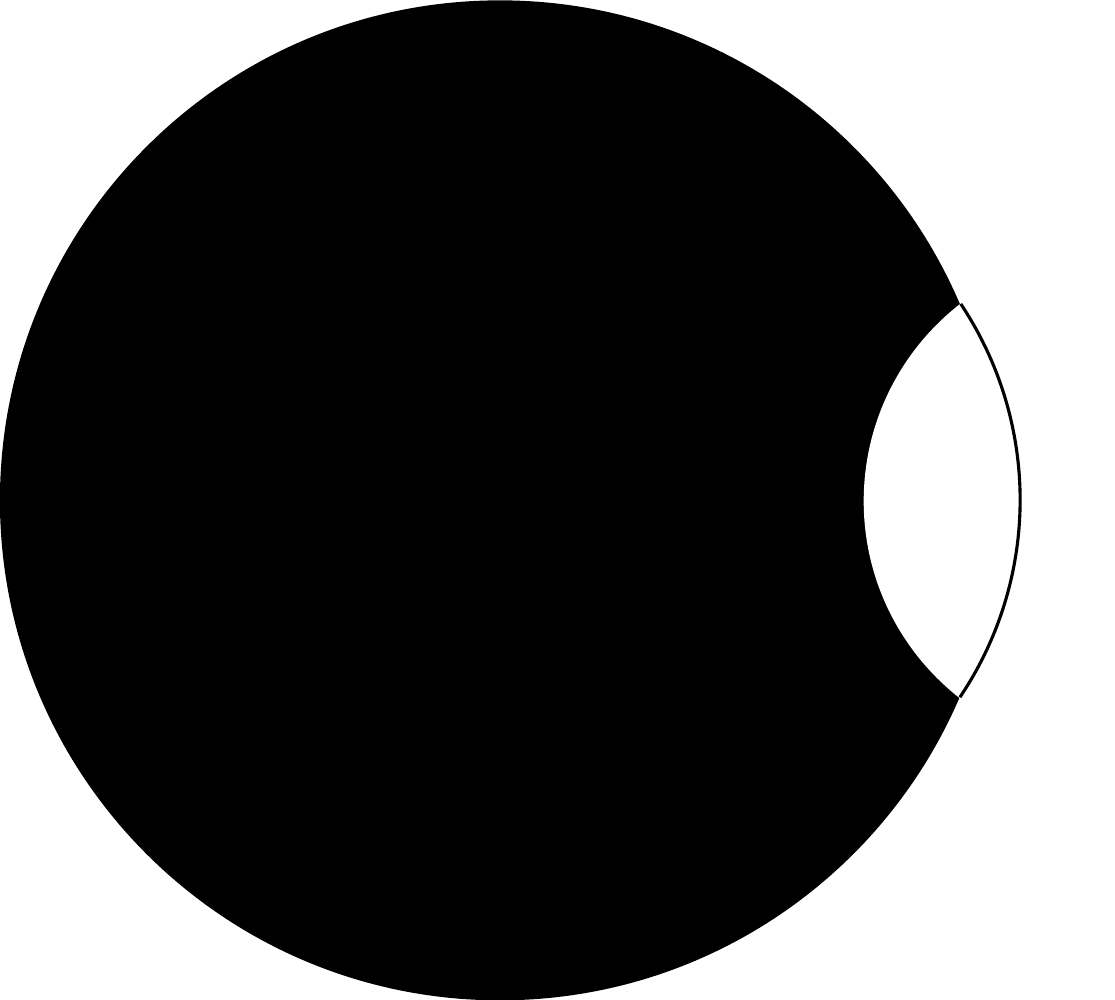}} & $(-,-,+,+)$ & $\sigma_{AB} \to \sigma_{AC} + \sigma_{BC}$\\
\hline

$(-,+,+)$ & \raisebox{-0.4in}[0.5in][0.5in]{\includegraphics[scale=0.2]{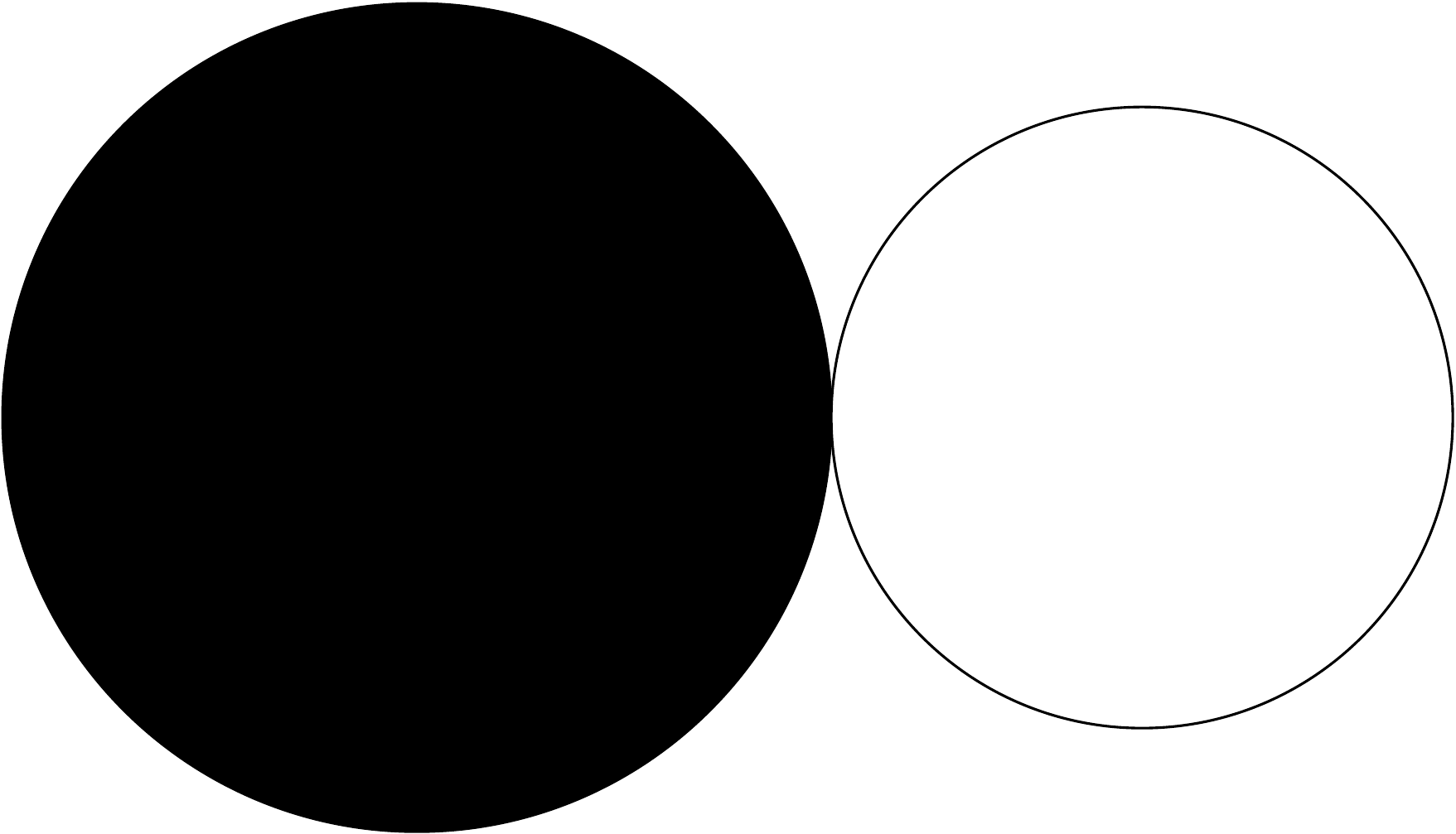}}  & \raisebox{-0.4in}[0.5in][0.5in]{\includegraphics[scale=0.2]{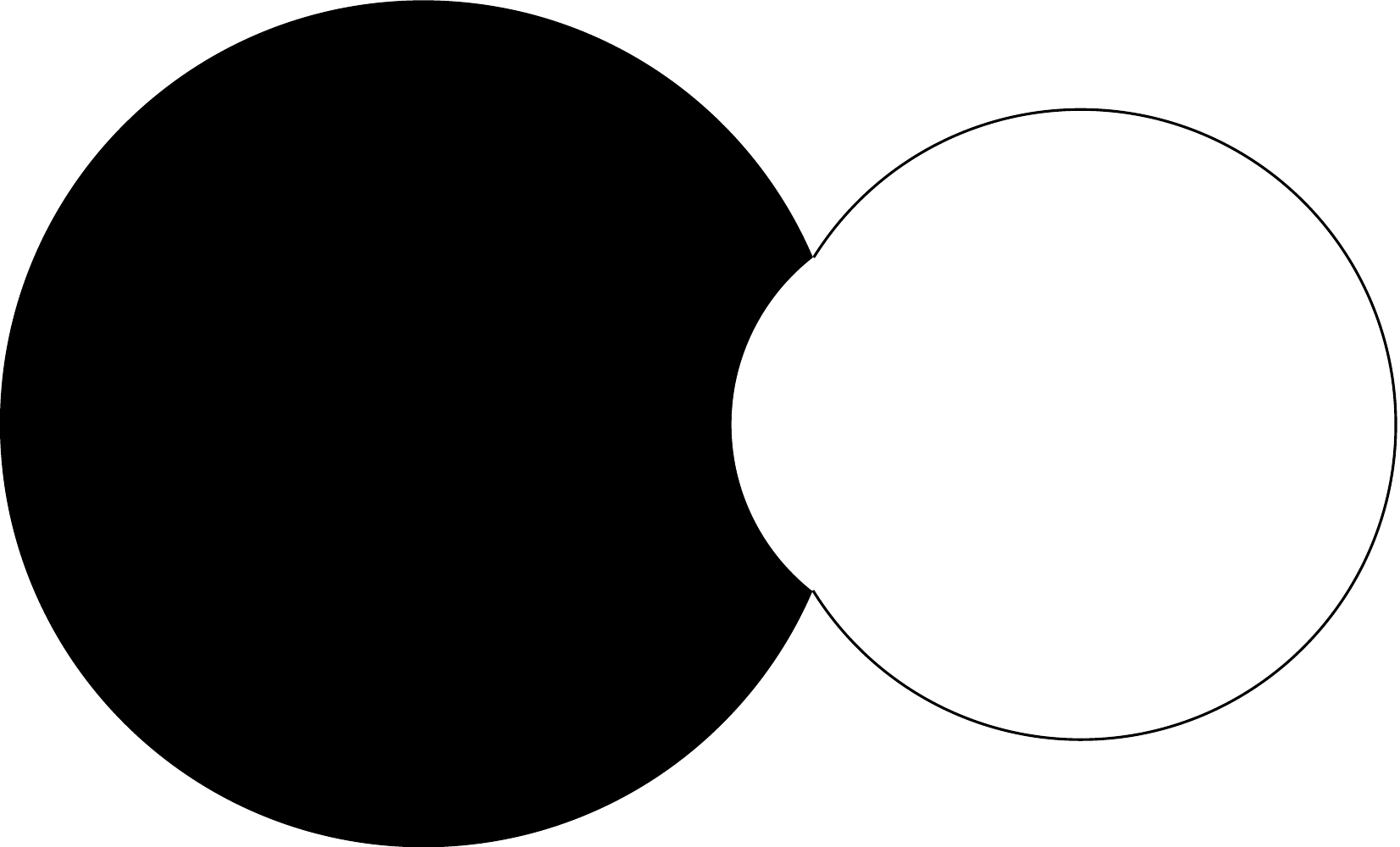}} &  $(+,-,-,+)$ &$\sigma_{BC} \to \sigma_{AB} + \sigma_{AC}$\\
\hline
$(-,+,-)$ & \raisebox{-0.4in}[0.5in][0.5in]{\includegraphics[scale=0.2]{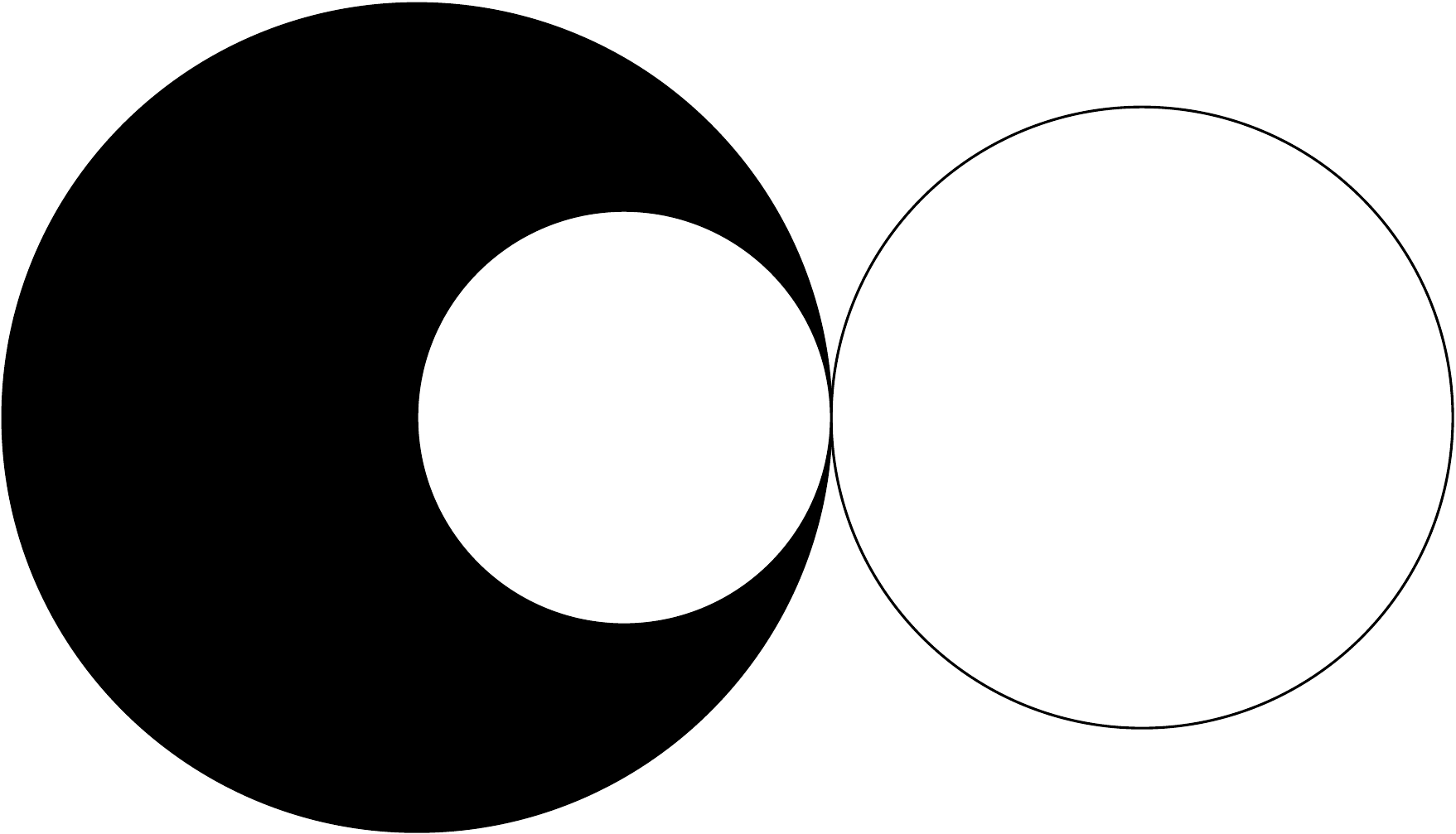}}  & \raisebox{-0.4in}[0.5in][0.5in]{\includegraphics[scale=0.2]{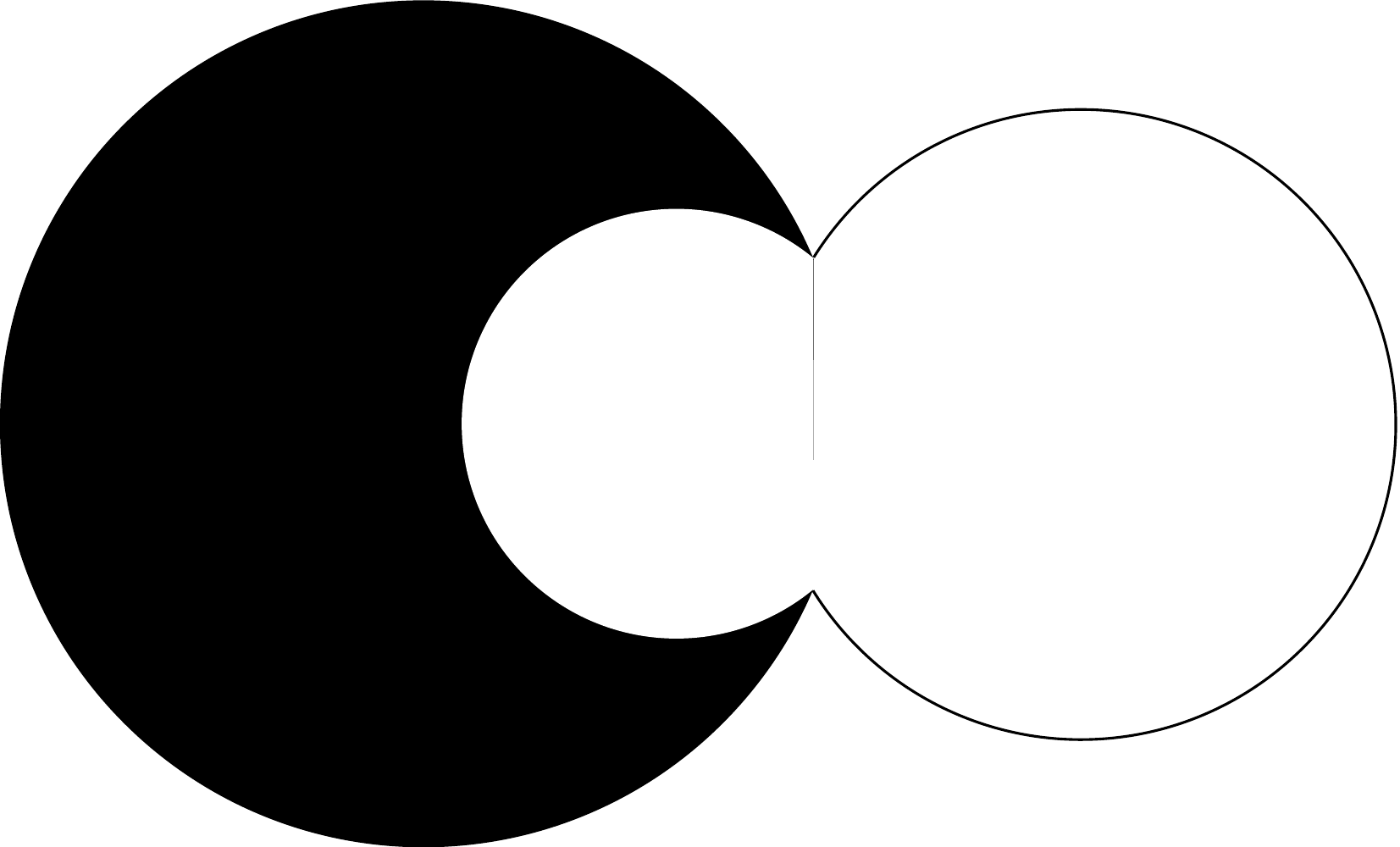}} & $\times$ & \parbox{1.4in}{\begin{center}does not exist \\ ~ \\ excluded by energy conservation\end{center}} \\
\hline
\end{tabular}
\caption{A catalogue of critical points of action (\ref{sfunctional}). The columns contain: (1) the signs of $(z_{AB}, z_{AC}, z_{BC})$; (2) cartoons of the $r=0$ saddle points; (3) cartoons of the $r=r_*$ saddle points; (4) the signs of the eigenvalues $(S_{rr}, S_{XX})$ at $r=r_*$ for $X=AB,AC,BC$; and (5) the regime in parameter space that enacts the limit $r_* \to 0$. This limit takes configurations from column (3) to those of column (2). The signs of the eigenvalues at the $r=0$ saddle points are the same as those listed in column (4) except $S_{rr}$ = 0, up to a positive junction correction. In all pictures, regions filled with the true vacuum C are drawn in white while regions filled with the intermediate vacuum B are filled with black.\newline}
\Label{allsols}
\end{table}

\section{Discussion}
\label{discussion}

For a tunneling event we want precisely one negative eigenvalue \cite{negeig, langer-decay}. The present work shows that in the thin wall approximation a multidimensional field landscape does not produce additional saddle points with one negative eigenvalue.  Although our analysis neglected the junction term $4\pi\kappa\mu^2r^2$, its inclusion does not affect this finding. 

To confirm this, consider the junction-deformed action 
\begin{equation}
S = S^{AB} + S^{AC} + S^{BC} + 4 \pi \kappa \mu^2 r^2 \,.\label{swithjunction}
\end{equation}
It contains at most four critical points. The first one, at $r=0$, is always present and has $S_{rr}=8\pi\kappa\mu^2>0$. The second one arises when the leading cubic term in the expansion (\ref{taylor}) of (\ref{sfunctional}) is negative: pitted against the positive quadratic term in (\ref{swithjunction}), it leads to an additional critical point with $S_{rr}<0$ and $r$ roughly of order $\kappa \mu^2 / \sigma$. In addition to these, there may be one or at most two other critical points, which reduce to $r=r_*$ as $\kappa \to 0$. Generically, a non-zero $\kappa$ affects the $r=r_*$ saddle point merely by a shift of $r_*$, without inducing any qualitative changes. Under some circumstances, however, the junction term can split the non-spherical saddle point
into a pair (one with positive and one with negative $S_{rr}$) and / or change the signs of $z_{AB}, z_{AC}, z_{BC}$, which affects the $S_{XX}$ eigenvalues. We have verified that all critical points with $r \neq 0$ have two or more negative eigenvalues.

The fact that in the thin wall approximation $S_{rr}$ at the dominant critical point is small (of order $\kappa \mu^2$) highlights an interesting possibility. As one varies the landscape parameters $\{\epsilon_{AB}, \epsilon_{AC}, \epsilon_{BC}, \sigma_{AB}, \sigma_{AC}, \sigma_{BC}\}$, one may bring the location of the non-trivial saddle point $r_*$ arbitrarily close to 0. From eq.~(\ref{masterr}) it is clear that this happens when the triangle with sides $\sigma_{AB},\sigma_{AC},\sigma_{BC}$ becomes degenerate. Physically, we recognize that in the strict triangle degeneration limits one may replace the heavier wall with the pair of lighter walls at no energetic cost. When this happens, the spherical bubbles (rows $(+,+,+)$ and $(-,-,+)$ in Table~\ref{allsols}) are destabilized: the maximum at $r=r_*$ merges with the extremum at $r=0$, which subsequently becomes a maximum of $S(r)$ so that the spherical instanton ceases to mediate vacuum decay. An example of this phenomenon was described in \cite{runaway}.

\section*{Acknowledgements}
We thank Alex Dahlen, I{\~n}aki Garc\'ia Etxebarria, Justin Khoury, Mark Van Raamsdonk, Moshe Rozali, Kris Sigurdson, and Eva Silverstein for discussions. VB is supported by DOE grant DE-FG02-95ER20893 and thanks the theory group at UBC for hospitality during this work. BC is supported by Natural Sciences and Engineering Research Council of Canada and thanks the University of Pennsylvania for hospitality during this work. KL and TSL are supported by Natural Sciences and Engineering Research Council of Canada and by the Institute of Particle Physics.

\end{document}